\documentclass[aps,twocolumn,superscriptaddress,showpacs,pre,nofootinbib,floatfix]{revtex4}
\usepackage{graphicx}

\begin{document}

\title{Improved analysis of SN1987A antineutrino events}

\author{G. Pagliaroli}
\email{giulia.pagliaroli@lngs.infn.it}
\affiliation{INFN, Laboratori Nazionali del Gran Sasso, Assergi (AQ), Italy}
\affiliation{ University of L'Aquila, Coppito (AQ), Italy}

\author{F. Vissani}
\email{francesco.vissani@lngs.infn.it}
\affiliation{INFN, Laboratori Nazionali del Gran Sasso, Assergi (AQ), Italy}

\author{M. L. Costantini}
\email{marialaura.costantini@lngs.infn.it}
\affiliation{INFN, Laboratori Nazionali del Gran Sasso, Assergi (AQ), Italy}
\affiliation{ University of L'Aquila, Coppito (AQ), Italy}

\author{A. Ianni}
\email{aldo.ianni@lngs.infn.it}
\affiliation{INFN, Laboratori Nazionali del Gran Sasso, Assergi (AQ), Italy}

\date{\today}

\begin{abstract}
We propose a new parameterization of the antineutrino
flux from core collapse supernovae, that allows an
interpretation of its astrophysical parameters within the
Bethe and Wilson scenario for the explosion, and
that leads to a reasonable (smooth) behavior of the average
energy and of the luminosity curve.
We apply it to analyze the events
observed by Kamiokande-II, IMB and Baksan detectors
in correlation with SN1987A.
For the first time, we consider
in the same analysis all data characteristics:
times, energies and angles of the observed events.
We account for the presence of background
and evaluate the impact of neutrino oscillations.
The hypothesis that the initial luminous phase of emission
(accretion) is absent can be rejected at the 2\% significance level.
Without  the need to impose external priors in the likelihood
analysis, the best-fit values of the astrophysical parameters are
found to be in remarkable agreement with the expectations of the
standard core-collapse scenario; in particular, the electron
antineutrino-sphere radius is 16 km, the duration of the accretion
phase is found to be  $0.55$ s, and the initial accreting mass is
0.22 $M_\odot$. Similarly the total energy emitted in neutrinos is
$2.2\times 10^{53}$ erg, again close to the expectations. The
errors on the parameters are evaluated and found to be relatively
large, consistently with the limited number of detected events;
the two dimensional confidence regions, that demonstrate the main
correlations between the parameters, are also given.

\end{abstract}

\maketitle
\section{Introduction}
We begin recalling the interest of core collapse supernovae,
the status of their understanding, and
the expectations for neutrino emission in the standard scenario.
Next, we discuss in Sect.~\ref{what}
the motivations for an improved analysis of SN1987A
observations in the context of the standard scenario.
Finally, we offer an outline of the present
investigation.



\subsection{Neutrino emission in core collapse supernovae\label{cont}}
Core collapse supernovae (SN) are astrophysical events
in which all known forces interplay with each other
in extreme physical conditions. An adequate
modeling of the processes occurring during this event
would be important to obtain information
on the left-over compact star  \cite{compactstar},
on nucleosynthesis \cite{remnants,nucleosint,Janka_2002ej},
on the properties of the supernova remnant \cite{snr},
and on the expected signals during the explosion;
in particular, gravitational waves and neutrinos~\cite{gw,Kotake_2005zn}.

Because of the complexity of the problem, the modeling of the
physical processes is still in evolution, but it is generally
accepted that the role of neutrinos is critical for the energy
transport as first suggested in \cite{clg}. The collapse and the
formation of a compact object, like a neutron star, have to pass
through substantial neutrino emission, see, e.g., \cite{burrows}.
The details of how the explosion takes place and how the neutrinos
are emitted are less clear and necessarily model dependent. In
this work we focus on the only mechanism that has been studied in
some detail: the {\em neutrino-driven mechanism} also known as
Bethe and Wilson scenario~\cite{bw} or delayed scenario for the
explosion. In the neutrino-driven mechanism, the explosion of the
massive star receives crucial assistance from the energy
deposition due to an initial, intense neutrino luminosity.
Although the viability of this mechanism cannot be considered
fully demonstrated at present\footnote{We note that relevant
discussion can be traced back to 1978 with the calculations of
Nadyozhin \cite{nad}.}, recent theoretical
results~\cite{Janka_2007yu,janka new} encourage the opinion that
the neutrino-driven mechanism works for certain core collapse~SN.

In the neutrino-driven mechanism, there are
two main phases of neutrino emission:\\
{\itshape{i}}) A thermal phase, called {\em cooling}, occurring
when the proto-neutron star cools quietly. This phase involves most
of the emitted neutrinos, 80-90\% in energy.\\
{\itshape{ii}}) A brief and very luminous neutrino emission,
here termed {\em accretion}, that should involve
a lower amount of neutrinos, 10-20\%
in energy.
The accretion phase characterizes the neutrino-driven mechanism of
the explosion,
and it is expected to occur in the first stage of neutrino emission. In
this phase, the matter is rapidly accreting over the proto-neutron star
through the stalled supernova shock wave.
The two most important processes of neutrino emission are
\begin{equation}
e^-\ p\to n\ \nu_e \mbox{ and } e^+\ n\to p\ \bar\nu_e \label{eq1}
\end{equation}
due to the abundant presence of nucleons and of quasi-thermal
$e^+e^-$ plasma. These types of neutrinos ($\nu_e$ and $\bar\nu_e$)
transfer to the star a small fraction of their energy, $f\sim 0.1$,
necessary
to revive the stalled shock wave.
See~\cite{janka} for a wide description
of the phase of accretion,  enriched by analytical arguments.

The neutrinos from phases
{\itshape{i}}) and {\itshape{ii}}) can
be observed in conventional supernova
neutrino detectors \cite{Burrows_1991} (namely, water Cherenkov
and scintillators). In particular electron antineutrinos give signal
mainly through inverse beta decay reaction on free protons:
\begin{equation}
\bar\nu_e \ p \to e^+ \ n
\label{ibd}
\end{equation}
Thus the existence of the accretion and cooling phases,
generically expected in the neutrino-driven scenario,
can be experimentally verified.


\subsection{What can we learn from SN1987A observations?\label{what}}
SN1987A is the first and the only occasion at present to test the
credibility of the various hypotheses on how a SN works. In fact,
the events observed by IMB~\cite{imb1,imb2}, Kamiokande-II
\cite{kam,kam2} and Baksan \cite{baksan}, represent a historic
opportunity to investigate the physics of the collapse and of the
explosion. A very extensive literature testifies the effort to
extract information from these data with a wide variety of methods
\cite{chud,daimo,preb,sat,jir,1stosc,
bahcall,bahcall05,bahcall1,bahcall2,bahcall22,
bahcall25,bahcall3,bahcall4,bahcall45, ll,
bahcall5,bahcall51,bahcall52, bahcall55, cc,rm,jcap,bahcall6}.
Usually, a specific characteristic of the SN1987A data is studied,
most frequently the energy distribution. The energy and the time
distributions are jointly considered in a few analyses, but most
often describing the neutrino emission with (overly) simple
models, a procedure that is only partially justified by the
limited amount of events collected.



The next neutrino observation will be an
extraordinary occasion to progress, but
recall that SN are rare on human time scale and it
is not possible to reliably predict when the next one will happen.
Thus, we should try our best using the only data that we have
at our disposal, and in particular, we should attempt to address
the question on
{\em whether there is a hint of accretion from SN1987A observations},
as expected. In this respect, a point that deserves to be stressed
is that all detectors observed a relatively large number
of events in the first second of data taking, about 40 \%:
there are in fact 6 events in Kamiokande-II,
3 events in IMB and 2 events in Baksan.

A milestone for the point of discussion
and more in general for SN1987A data analysis
is the paper of Lamb and Loredo~\cite{ll} (LL in the following), where
it is argued that the SN1987A
observations can be used to claim for
an evidence of the accretion phase.
The LL paper, widely cited in theoretical and experimental reviews, is
generally considered a useful application of refined statistical
techniques and, in the present paper, we will provide the first independent
verification of their results.
However, we deem that, in view
of the importance of their claim and in the light of various advances in
neutrino physics (e.g., in oscillations,~\cite{msw1,msw2,msw3,msw4} and
\cite{sk,sno,kl,k2k,minos}),
it is necessary to offer a critical
discussion of the assumptions of the analysis by Lamb and Loredo.
More specifically:\\
1) the likelihood structure can be enhanced including a more accurate
detection cross section~\cite{sss}, the information on the directions
of the events,
a different treatment of the background~\cite{jcap};\\
2) the theoretical model for neutrino emission, required for
the data analysis, can be improved and
approached as much as possible to the real signal
expected from the numerical simulations of neutrino
emission.
In this respect, a specific criticism was raised by Raffelt and
Mirizzi~\cite{rm}, who emphasized that the two-phase
parameterization used in LL analysis leads to
a sudden jump of the average neutrino energy while passing
from accretion to cooling. This behavior is rather different
from the expectations of the numerical simulations.

In short, our main tasks are to present a somewhat different likelihood,
to propose  an improved parameterization for neutrino flux,
to include oscillations,
and finally to evaluate the impact of the various
new points for the analysis of SN1987A observations.

\subsection{Layout of the work}
The structure of this paper is the following: in Sect. \ref{likimpr}
we discuss
the construction of the likelihood function involved in the
analysis of the data set, underlining the improvements carried
in the description of detection rate; in Sect.~\ref{sec:m} we
propose a new parameterization for neutrino emission, building it step by step
and  discussing its features, leaving the description of
certain technical details to the appendix; finally in
Sect.~\ref{sec:r} we draw the results of our analysis.

\section{Likelihood construction\label{sec:l}}
\label{likimpr} In this section we describe the likelihood that we
adopted to compare the observed events and the assumed
flux for the neutrino emission,
stressing the novelties and the technical improvements.

\subsection{Signal rate}\label{a}
The signal in each detector, triply differential in time, positron energy $E_e$ and cosine of the angle $\theta$ between the antineutrino and the
positron is:
\begin{equation}
\begin{array}{r}
\displaystyle R(t,E_e,\cos\theta)= N_p \frac{d\sigma_{\bar\nu_e p}}{d\cos\theta}(E_{\nu},\cos\theta )\
\Phi_{\bar\nu_e}(t, E_{\nu}) \times\\
\displaystyle \times \xi_d(\cos\theta)\ \eta_d(E_e)\ \frac{dE_{\nu}}{dE_e},
\label{seta}
\end{array}
\end{equation}
where $N_p$ is the number of targets (=free protons) in the detectors,
$\sigma_{\bar\nu_e p}$ is the inverse beta decay
cross section (Eq.~\ref{ibd}), $\eta_d$ the--detector dependent--average detection
efficiency, $\xi_d$ is the angular bias =1 for Kamiokande-II and Baksan whereas
for IMB $\xi_d(\cos\theta)=1+0.1 \cos\theta$~\cite{imb2}, and,
finally,
$\Phi_{\bar{\nu}_e}$ is the electron antineutrino flux, differential in
the antineutrino energy $E_\nu$ and discussed later in this work.

The expected number of signal events $\mu_s$ is the crucial ingredient,
along with the expected number of background events
$\mu_b$, to construct the Poisson likelihood $\mu^n \exp(-\mu)/n!$,
where $\mu=\mu_s+\mu_b$ and where $n$ is the
number of observed events.
To  evaluate $\mu_s$ we use Eq.~\ref{seta}:
the number of expected signals in a bin
is just $R(t,E_e,\cos\theta) dt dE_e d\cos\theta$;
the total number of the
events is the integral of $R(t,E_e,\cos\theta)$ over its
three variables.

As mentioned in the introduction,
our analysis is similar to the one of
Lamb and Loredo \cite{ll}, with whom we agree within
errors when we strictly stick to their procedure.
The signal rate that we adopted in this paper departs
from their one in the following points:

\subsubsection{Cross section and event direction\label{xsec}}
We adopt the inverse beta decay cross section calculated in \cite{sss} and in particular we use the differential expression $d\sigma_{\bar\nu_e
p}/d\cos\theta$ given in Eq.~(20) of that
paper. The energy of antineutrino is given
in terms of the positron energy $E_e$ and of the angle $\theta$
between the antineutrino and the positron directions:
\begin{equation}
E_\nu=\frac{E_e+\delta_{-}}{1-(E_e-p_e\cos\theta)/{m_p}},
\label{kinem}
\end{equation}
where $\delta_{-}=(m_n^2-m_p^2-m_e^2)/(2 m_p)= 1.294\mbox{ MeV}$
and $p_e$ is the positron momentum.
The new total cross section agrees at 10 MeV with the one used by
Lamb and Loredo, whereas at 20 MeV (30 MeV) it is 6\% (12\%)
smaller.

\subsubsection{Efficiency\label{eff}}
Following the traditional approach and
differently from~\cite{ll} we include in Eq.~\ref{seta}
the detection efficiency as a function of the true energy of
the event. A formal justification of our procedure is
given in \cite{veno}, that is in contrast
with the formal justification in
Appendix A of \cite{ll}.

Our procedure simply accounts for the
evident fact that the expected
number of {\em signal} events $\mu_s$ should include all relevant
detector dependent features: loss of events due to light attenuation,
fluctuations of the number of photoelectrons,
detector geometry, {\em etc.}. These features produce an imperfect
($\eta_d<100\%$) detection efficiency, that means that only a fraction
of the produced positrons is actually detected.
We have in mind an {\em average efficiency}
evaluated by a MC procedure, namely
1)~simulating several events
with true energy $E_e$ but located in the
various positions and emitted in all possible directions, then
2)~counting the fraction of times that an event is recorded, finally
3)~deducing also the `smearing' (=average error as a
function of~$E_e$).

It is possible to argue in favor
of our procedure by considering
the following situation. Imagine two detectors that differ
by the detector efficiency: $\eta=100\%$
in the first one, $\eta=10\%$ in the second, and with a
rate of positron production equal to the background rate.
Suppose that each of
these detectors observed one event. According to the procedure of
\cite{ll} (used,  e.g., to obtain their Tab.~VI) the probability that the
observed event is due to a signal is 50\% in both detectors, that we
find paradoxical: the better the detector, the higher should be the
chances of a signal. Instead, adopting our procedure,
the probability that the event is due to
a signal is 50\% in the first detector and 9\% in the second
one, which we find more plausible.

We recall (in agreement with \cite{ll,jcap})
that for an even more refined analysis of the data,
one should not use the average detection efficiency, but should
rather evaluate the specific detection efficiency and
background rate for any individual event. In our understanding, a
correction on individual basis of this type
was performed only to assess the errors on the
energies of the events, see~\cite{kam2}.

\subsection{The assumed likelihood}
We estimate the theoretical parameters by the $\chi^2$:
\begin{equation}
\chi^2\equiv-2 \sum_{d=k,i,b} \log( {\cal L}_d ),
\end{equation}
where  ${\cal L}_d$ is the likelihood of any detector ($k,i,b$ are shorthands for Kamiokande-II, IMB, Baksan). We use Poisson statistics; dropping
constant (irrelevant) factors, the `unbinned' likelihood of each of the 3 detectors is:
\begin{equation}
\begin{array}{l}
{\cal L}_d= e^{-f_d \int \!\!  R(t) dt} \times \prod^{N_d}_{i=1} e^{R(t_i) \tau_d  }\times\\[1ex]
\times \left[ \frac{B_i}{2} +\! \int\!\! R(t_i, E_e,
\cos\theta_i) {\cal L}_i(E_e ) dE_e \right]. \label{liky}
\end{array}
\end{equation}
We denote by $R(t)$ the integral of $R(t,E_e,\cos\theta)$ over the variables $E_e$ and $\cos\theta$. In IMB, the live-time fraction is $f_d=0.9055$ and
the dead-time is $\tau_d=0.035$~s, whereas for the other detectors $f_d=1$ and $\tau_d=0.$ Each detector saw $N_d$ events; their time, energy and
cosine with supernova direction are called $t_i$, $E_i$ and $c_i$ ($i=1...N_d$).

${\cal L}_i$ is a Gaussian distribution that
includes the estimated values of
the energy $E_i$ and the error of the energy $\delta E_i$ for each individual
event, accounting for the detector-dependent effects
on energy measurements.
The inclusion of the error
on the measurement of $\cos\theta$ does not change
significantly the likelihood, so we simply set $\cos\theta=c_i$ for
each event.\footnote{For the first 12 Kamiokande-II and for the 8 IMB events
the value of $\cos\theta=c_i$ is measured; we set
instead $c_i=0$ for the 5 events of Baksan and the
last 4 events (out of 16) of Kamiokande-II.}
Finally, we do not include an
error on the event times
$\delta t_i$, since the relative time of each event is precisely measured.

The absolute times have not been measured precisely enough
(except for IMB). However, the procedure of analysis that we adopt
requires only the relative times between the events:
the experimental input is
$\delta t_{i}=t_i^{\mbox{\tiny exp}}-t_1^{\mbox{\tiny exp}}$.
The times $t_i$ are defined as follows:
\begin{equation}
t_i=t^{\mbox{\tiny off}}+\delta t_{i}
\end{equation}
where $t^{\mbox{\tiny off}}\ge 0$ is the offset (or
delay) time between the first neutrino that reached the Earth (that, by definition, occurred at $t=0$) and the first event that was detected. We
introduce one parameter $t^{\mbox{\tiny off}}$ for each detector, and fit
their values from the
data.
The integral over the time in the
first exponential factor of Eq.~\ref{liky}
is performed from the moment when the first neutrino reaches the Earth till the end of data taking, $t=30$~s; the condition
that all the data are included imposes mild restrictions, such as
$t^{\mbox{\tiny off}}_{\mbox{\tiny KII}}<6$ s, that do  not
have a relevant role in the
analysis.

\subsubsection{Background\label{bck}}

The probability that an event is due to background is denoted by
$B_i$ in Eq.~\ref{liky}. It is calculated as $B_i=B(E_i)$: $B_i$
is the {\em measured} background rate for the given energy
[Hz/MeV]. The background distribution differential in time, energy
and cosine is $B(E_e)/2$; the factor $1/2$ describes a uniform
cosine distribution. This definition is different from the one of
LL, $B_i=\int\!\!B(E_e) {\cal L}_i(E_e ) dE_e$, that
has been argued to be inaccurate in \cite{jcap}.
The values of $B_i$ that we use for Kamiokande-II
are given in Appendix A of \cite{jcap}.
The events of Kamiokande below 7 MeV have a higher
background rate than found by LL, those above 9 MeV a lower
background rate, while the other ones stay almost unchanged.
The changes for Baksan are instead negligible. It is fair to assume in
good approximation that, in the time window of interest, IMB
observations are safe against background contamination.

\section{Antineutrino flux description}\label{sec:m}

In this section we describe the
parameterization of the neutrino flux that we adopt.
We describe the signal introducing three `microscopic'
(i.e., physically meaningful) parameters
for each phase, that, roughly speaking, are needed
to quantify the duration of the emission process,
the intensity of the emission and the
average energy of the antineutrinos.
The adopted time distributions are constructed to enforce
the continuity of the instantaneous luminosity
and of the average energy  as found in the numerical simulations.
Moreover, we tried to maintain the parameterization as simple as possible.


\subsection{ Parameterized antineutrino fluxes\label{fluka}}
\subsubsection{Cooling phase}

In the last phase of the SN collapse the nascent proto-neutron star
evolves in a hot neutron star (with radius $R_{ns}$) and this process is
characterized by a neutrinos
and antineutrinos flux of all species.
This is the  \emph{cooling} phase;
we use the suffix {\em c} in the corresponding symbols.

A rather conventional parameterization of the electron
antineutrino flux, differential in the energy is:
\begin{equation}
\Phi^0_{c}(t,E_\nu)= \frac{1}{4\pi D^2} \frac{\pi c}{(h c)^3}\left[
 4\pi R^2_c\ g_{\bar\nu_e}(E_\nu,T_c(t))\right] \label{fluso}
\end{equation}
where the Fermi-Dirac spectrum of the antineutrinos is
\begin{equation}
g_{\bar\nu_e}(E_\nu,T_c(t))=\frac{E_\nu^2}{1+\exp[E_\nu/T_c(t)]}
\end{equation}
The time scale of the process is included in the function:
\begin{equation}
T_c(t)=T_c \exp[{-t/(4\tau_c)}].
\end{equation}
Eq.~\ref{fluso} describes an isotropic emission of
antineutrinos from a distance  $D (=50$~kpc in the case of SN1987A).
[We use the symbol $\Phi^0$ rather than $\Phi$ to
emphasize that flavor oscillations have not been included yet].

The astrophysical free parameters
are $R_c$, $T_c$, and $\tau_c$ namely:
the
radius of the emitting region (neutrino sphere),
the initial temperature,  and the time
constant of the process. We recall which are the generic expectations:
$R_c\sim R_{ns}=10-20$~km, $T_c=3-6$~MeV, and $\tau_c=$few-many seconds.
Rather than using these {\em a priori}, we will deduce the value of these
parameters by fitting the SN1987A data,
and later, we will compare the results with the expectations.

\subsubsection{Accretion phase\label{am}}
After the bounce, the simulations indicate that the
shock wave, propagating into the outer core of the star,
looses energy and eventually gets stalled. It forms
an accreting shock that encloses a region of dissociated
matter and hot $e^+ e^-$ plasma, where the weak reactions
of Eq.~\ref{eq1} give rise to intense
$\nu_e$ and $\bar\nu_e$ luminosities.
This emission lasts a fraction of a second.
In the Appendix we describe in more details the
conceptual scheme for $\bar\nu_e$ emission:
a neutron target exposed to a flux of thermal positrons.
The neutrons are treated as a transparent target, for
only a small fraction of antineutrinos is expected to couple with the star.
This is the  \emph{accretion} phase;
we use the suffix {\em a} in the corresponding symbols.

The parameterized $\bar\nu_e$ flux is
\begin{equation}
\begin{array}{l}
\Phi^{0}_{a}(t,E_\nu)= \frac{1}{4\pi D^2} \frac{8\pi c}{(h c)^3}\\
\times \left[N_n(t)\sigma_{e^+n}(E_\nu) \ g_{e^+}(\bar{E}_{e^+}(E_\nu),T_a(t))\
\right],
\end{array}
\label{flusso}
\end{equation}
where $N_n(t)$ is the number of target neutrons
assumed to be at rest and the thermal flux of positrons:
\begin{equation}
g_{e^+}(E_{e^+},T_a(t))=
\frac{E_{e^+}^2}{1+\exp \left[ E_{e^+}/T_a(t) \right]}
 \label{fluacc}
\end{equation}
is calculated at an average positron energy, namely
$\bar{E}_{e^+}(E_\nu)=\frac{E_\nu-1.293\rm MeV}{1-E_\nu/m_n}$.
In the energy range of interest, $E_\nu=5-40$ MeV, a simple numerical
approximation of the cross section for positron interactions is
\begin{equation}
\sigma_{e^+n}(E_\nu)\approx \frac{4.8\times 10^{-44}
E_\nu^2}{1+E_\nu/(260\mbox{ MeV})}
\label{appra}
\end{equation}
The derivation of Eqs.~\ref{flusso} and \ref{appra}
is given in the appendix.

The average energy of the antineutrinos
is roughly given by $5\, T_a$
and the spectrum is slightly non-thermal,\footnote{The
deviation from the thermal distribution
can be described by a `pinching factor'--an {\em effective} chemical
potential introduced to distort
the Fermi-Dirac thermal
spectrum--in the range 4-5, that decreases when $T_a$ increases;
e.g., for $\delta E_\nu/T_a=0.41$ the pinching factor is 4.2.}
mostly due
to the presence of the cross section $\sigma_{e^+n}$.
For example, when $T_a=1.5,\ 2.5,\ 3.5$ MeV, we get
$\langle E_\nu \rangle/T_a=5.5,\ 5.2,\ 5.0$ respectively and
$\delta E_\nu/T_a=0.39,\ 0.41,\ 0.41$ where $\delta E_\nu\equiv\sqrt{
\langle E_\nu^2 \rangle-\langle E_\nu \rangle^2}$.
Another manifestation that the distribution is
non-thermal is the scaling of the luminosity with the temperature,
roughly  as $T_a^6$--different from the thermal scaling $T_a^4$.
In order to give some feeling of the
antineutrino emission, if we suppose that
at $t=0$ we have
$T_a=2.5$~MeV and $M_a=0.15\ M_\odot$ (see just below
for a precise definition of these quantities)
we get a luminosity of $1.1\times 10^{53}$
erg/s; the same luminosity and average energy
would be given by a black body  distribution with
$T_c=4.1$~MeV and $R_c=82$~km.

There are two time dependent quantities in Eq.~\ref{flusso}:
the number of neutrons $N_n(t)$ and the positron temperature $T_a(t)$.
It is straightforward to introduce a temperature that interpolates
from an initial value to a final value:
\begin{equation}
T_a(t)=T_i + (T_f-T_i) \left( \frac{t}{\tau_a} \right)^m
\!\! \mbox{ with }
\left\{
\begin{array}{l}
T_i=T_a \\
T_f=0.6\ T_c
\end{array}
\right.
\label{tap}
\end{equation}
where $T_a$ denotes the {\em positron} temperature
at the beginning of accretion (to be contrasted with
$T_c$, the {\em antineutrino} temperature
at the beginning of the cooling phase).
With this parametrization, the positron temperature reaches
$0.6\ T_c$ at $t=\tau_a$, that is what is needed match the average
antineutrino energies, namely, to ensure a continuous behavior of the
average antineutrino energy (in particular  at the end of the accretion phase
and at the beginning of the cooling phase).
The power $m=1-2$ mimics the behavior found in numerical simulations;
we adopt $m=2$ as a default value.

Now we discuss the time evolution of the number of neutrons
exposed to positrons $N_n(t)$, proportional to
the luminosity in accretion. Our goal
would be a luminosity that, at least for $t\sim 0$,
decreases as $1/(1+t/0.5\mbox{ s})$.
This behavior is  suggested by the numerical simulations
and is advocated by LL \cite{ll}.
However, when we allow the temperature to vary,
we vary also the luminosity, that scales as
$N_n T_a^6$. Thus, we need to include an
explicit factor $(T_a(t)/T_a)^6$.
We arrive at our prescription for
the number of neutrons exposed to thermal positrons:
\begin{equation}
N_n(t)=\frac{Y_n}{m_n}\times M_a\times \left( \frac{T_a}{T_a(t)}
\right)^6
\times \frac{j_{k}(t)}{1+t/0.5\mbox{ s}}, \label{nnbest}
\end{equation}
the fraction of neutrons being set to $Y_n=0.6$.
$M_a$ is the initial accreting mass exposed
to the positrons thermal flux.
The time-dependent factor
\begin{equation}
j_k(t)=\exp[-(t/\tau_a)^k], \label{jcut}
\end{equation}
is included to terminate the
accretion phase at $t\sim \tau_a$.
LL use $k=10$, which however leads to a very sharp drop of the luminosity
at $t\sim \tau_a$.
In our calculations,
we will set instead $k=2$, a choice that offers
the advantage of leading to a smooth
(reasonable, continuous) luminosity curve, closer
to the type of curves found in numerical simulations.
[We will show in the next section
luminosity curves with $k=10$ and with $k=2$.]

Ultimately, the accretion phase involves 3 free parameters:
$M_a$, $T_a$ and $\tau_a$, the same number of
parameters of the cooling phase. Finally, we give rather generic
expectations on values of these parameters: $M_a$ is certainly lower
than the whole outer core mass (about $0.6\ M_\odot$); $T_a$ is expected to sit
in the few MeV range; and, finally, the accretion should last
a fraction of a second, $\tau_a\sim 0.5$~s being a typical number.

\subsection{Temporal shift}\label{shift}
In the model advocated by Lamb and Loredo and adopted for the
analysis of SN1987A data, the accretion and the cooling phases are
contemporaneous. This has the consequence that for times
$t<\tau_a$, the antineutrino distribution is not a thermal
spectrum, but a composition (=sum) of two thermal spectra, whose
average energies differ by more than a factor of 2 in the best-fit
point. At low energies, the spectrum is dominated by the
antineutrinos from accretion; at the highest energies, by the
antineutrinos from cooling. The possibility to have a {\em
composite spectrum} implies, among the other things, that it is
easy (trivial) to reconcile the low energy events observed by
Kamiokande-II and the high energy events observed by IMB in the
first second: they simply belong to different and simultaneous
phases of emission. We will verify this statement later by a
straightforward calculation--see Tab.~\ref{tabp2}. We note in
passing that such a composite spectrum is used in the work of
Loredo and Lamb \cite{ll} but this characteristic feature is
neither commented or discussed there. The compositeness of the LL
spectrum (i.e., the non-thermal tail) can be better perceived
plotting it in logarithmic scale and/or by considering the
time-integrated spectrum, see Fig.~2 of \cite{pag}.

However, at the best of our knowledge there is no evidence from
numerical simulations of a composite behavior of the spectrum; the
instantaneous $\bar\nu_e$ spectrum is found to be quasi-thermal at
any time and typically this property is shared also by the
time-integrated spectrum; see, e.g., the discussions in
\cite{dighe,cphd,nadr}. Deviations from a thermal distribution are
observed, especially during accretion \cite{nado,bahcall1,keil};
they can be effectively described  by a `pinching' parameter of
the order of a few, that means that the high-energy tail of the
spectrum is depleted--not enhanced as for the composite model
advocated by LL.

In short, we believe that, in absence of an explicit indication
from numerical simulations, the model that should be adopted in
data analyses (and/or the null-hypothesis that should be tested)
is the simplest one compatible with the numerical simulations,
namely: a $\bar\nu_e$ spectrum quasi-thermal at any time--rather
than, e.g., `composite' \cite{ll}, bimodal
\cite{bahcall6}
or exponentially decreasing in the energy \cite{rm}.
For this reason, we parameterize
the antineutrino flux as follows:
\begin{equation}
\Phi_{\bar\nu_e}(t) = \Phi_{a}(t) + (1-j_k(t))\times \Phi_{c} (t-\tau_a).
\end{equation}
where $\Phi_a$ (resp., $\Phi_c$) is given in
Eq.~\ref{flusso} (resp., Eq.~\ref{fluso}) in the
case when oscillations are absent.
Recalling that $j_k(t)$ appears explicitly in Eq.~\ref{nnbest}, i.e.,
in the accretion flux, it is clear that the effect of this function
is simply to interpolate between the two phases
of neutrino emission; in other words, the cooling phase begins around
$t=\tau_a$.

We note here a limit in which the likelihood becomes unphysical.
Consider the case when only the first
event detected by Kamiokande-II
falls in the accretion phase.
If the accretion were to last a very short amount of time,
$\tau_a \ll \delta t_2$, and if
$t^{\mbox{\tiny off}}\ll \delta t_2=0.107$~s,
the antineutrino flux at the time of the first event $t=\delta t_1=0$
could become very large even if the number of expected number of events
remains small. In this limit, the interaction rate in Eq.~\ref{seta}
becomes large, too. Thus, due to the factor $R(t_1,E_e,\cos\theta_1)$
in Eq.~\ref{liky}, the likelihood can be made arbitrarily large,
because the exponential factor in the same equation
(that depends on the expected number of events) will not change much.
The way to avoid this pitfall is to require a lower limit on $\tau_a$
in the numerical calculations. A limit that is adequate for the analysis of
data from SN1987A is $\tau_a>0.3$ s. We will demonstrate explicitly
the presence of this unphysical behavior
of the likelihood function later, when discussing its maxima.

\subsection{Neutrino oscillations}\label{osc}
Here we discuss the effects of neutrino oscillations on the
observed $\bar\nu_e$ fluxes. The survival probability $P$ for
$\bar{\nu}_e$ emitted by supernova is dictated by two different
interactions with the star medium. The first is the usual matter
effect \cite{dighe,fogli} of charged current interactions between
$\bar\nu_e$ and the electrons of the matter. The second is the
effect of $\nu-\nu$
interactions~\cite{nunu1,nunu2,nunu3,nunu4,nunu5}, that is known
to be important in specific cases \cite{somec1,somec2} and whose
behavior has been quantified in certain
approximations~\cite{simpl1,simpl2,simpl3,simpl4}. In order to
describe oscillations we have to distinguish the two arrangements
of the neutrino mass spectrum that are compatible with the present
knowledge of neutrino properties (see, e.g., \cite{revia}): For
{\bf normal} mass hierarchy the survival probability and the
observed  $\bar\nu_e$ flux are:
\begin{equation}
\begin{array}{ll}
P=U_{e1}^2 ,\\[1ex]
\Phi_{\bar\nu_e}= P\ \Phi_{\bar\nu_e}^0 + (1-P)\ \Phi_{\bar\nu_\mu}^0,\\
\end{array} \label{normal}
\end{equation}
where we recall that $\Phi^0$ is the flux in absence of
oscillations. We have assumed that
$\Phi_{\bar\nu_\mu}^0=\Phi_{\bar\nu_\tau}^0$ and each term is the
sum of the cooling flux and accretion flux. The $\nu-\nu$
interaction is most relevant for {\bf inverted} mass
hierarchy~\cite{simpl1,simpl2,simpl3,simpl4,simpl5,simpl6}. As an
example the electron antineutrino survival probability given in
\cite{simpl4} is:
\begin{equation}
\begin{array}{ll}
P=U_{e1}^2 (1- P_f) + U_{e3}^2 P_f,\\[1ex]
\Phi_{\bar\nu_e}= P\ \Phi_{\bar\nu_e}^0 + (1-P)\ \Phi_{\bar\nu_{\mu}}^0.\\
\end{array} \label{inverted}
\end{equation}
We adopt the usual decomposition of the mixing elements in terms of the mixing  angles:
$U_{e3}=\sin\theta_{13}$ and
$U_{e1}=\cos\theta_{12}\cos\theta_{13}$ with
$\theta_{12}=35^\circ\pm 4^\circ$ and
$\theta_{13}<10^\circ$ at 99 \% C.L.
For the measured solar oscillation parameters, the Earth matter
effect is known to be small (see, e.g.,~\cite{cavanna}).
We include it in the analysis anyway evaluating
the survival probabilities with
the PREM model~\cite{prem}. \\
In the case of
normal mass hierarchy the probability
$P\sim 0.7$ is reliably predicted and rather precisely
known. Instead, for inverted mass hierarchy, $P$
depends strongly on the unknown mixing angle $\theta_{13}$.
In fact, the flip probability $P_f$
(that quantifies the loss of adiabaticity
at the `resonance' related to the atmospheric $\Delta m^2$) is:
\begin{equation}
P_f(E_\nu,\theta_{13})=\exp\left[-\frac{U_{e3}^2}{3.5\times 10^{-5}}
 \left( \frac{20\mbox{ MeV}}{E_\nu}\right)^{2/3}  \right], \label{peffe}
\end{equation}
where the numerical value corresponds to the
supernova profile $N_e\sim 1/r^3$ given in \cite{fogli}.
The predictions for the $\bar{\nu}_e$ flux in
the detector could be tested with a relatively large amount of data:
a galactic supernova could turn out to be useful
to discriminate the right mass hierarchy.


We assume that an equal amount of energy goes
in each species (equipartition hypothesis)
during cooling.
Furthermore, we suppose
that the temperature of $\bar{\nu}_\mu$ and $\bar{\nu}_\tau$
is in a fixed ratio with the $\bar{\nu}_e$ temperature.
Following~\cite{keil}
we assume:
\begin{equation}
T(\bar\nu_\tau)/T(\bar\nu_e)=T(\bar\nu_\mu)/T(\bar\nu_e)=1.2, \label{defa}
\end{equation}
We tested that a value in the 1.0-1.5 or a deviation of the amount
of energy stored in non-electronic neutrino species by a factor of
2 does not affect crucially the fitted antineutrino flux. In the
accretion phase,  we will suppose that only $\nu_e$ and
$\bar{\nu}_e$ are emitted in equal amount, whereas
$\Phi^{0}_{a}(\bar\nu_{x})=0$. The fit provides a reasonably fair
description of the antineutrino flux anyway, but the estimation of
the amount of energy emitted during accretion should be considered
as a lower bound.

\section{Results\label{sec:r}}
In the previous two sections we constructed, step by step, a
likelihood function that represents the probability function of
the overall data set. This probability varies in the parameters
space and depends on the model for the antineutrino emission. As
anticipated, the main goal of this paper is to verify if the
best-fit values of the parameters, that maximize the likelihood
function, are physically acceptable.

We remark that LL adopted a Bayesian analysis while we use a
frequentist approach, which occasionally leads to some difference
in the confidence levels, but not in the best fit points. More
specifically to calculate the error on one parameter we use as a
rule the profile likelihood, namely the likelihood evaluated
fixing the parameter of interest and maximizing the other
(nuisance) parameters. A similar procedure allows us to calculate
the confidence regions.

The structure of this section is the
following: in Sect.~\ref{1cm} we describe the simpler (one
component)
model, in Sect.~\ref{2cm} we describe the more complete (two
components) model. This second part illustrates the impact
of each improvement in the description of the flux;
the final result is discussed in detail in Sect.~\ref{summat}.

\subsection{One component model\label{1cm}}

\begin{figure}[t]
$$\includegraphics[width=0.48\textwidth,
height=0.41\textwidth]{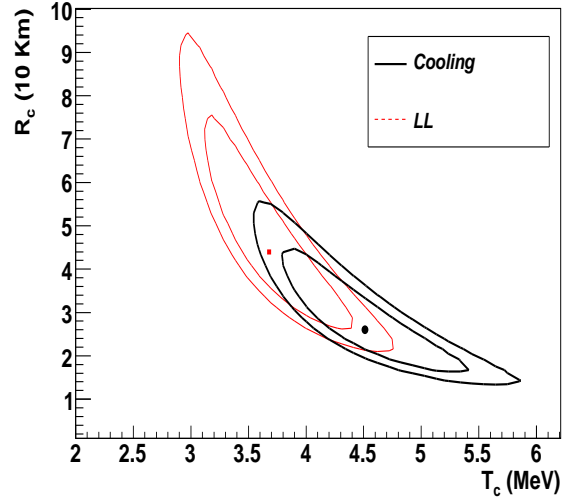}$$ \vskip-4mm \caption{\em
Two dimensional confidence regions for the cooling parameters
$R_c$ and $T_c$. The gray (red) contours are the exponential
cooling result ($68\%$ and $90\%$ C.L.) following Loredo and Lamb
analysis. The dark contours are our result for one component
cooling model. The best fit points are also displayed.
\label{fig0}}
\end{figure}

Here we list the results of the likelihood maximization procedure
when we include only the cooling phase. This model has 6
parameters: the $\bar{\nu}_e$ temperature $T_c$, the duration
$\tau_c$, the neutrinosphere radius $R_c$ and the three detector
offset times $t^{\mbox{\tiny off}}$. It is identical to the
``Exponential cooling model'' reported in \cite{ll}, also termed
``minimum'' or ``standard'' model in \cite{bahcall}. The
comparison of the best-fit values  of LL ($R_c=40$ km, $T_c=3.81$
MeV, $\tau_c=4.37$ s and $t^{\mbox{\tiny off}}=0$ s \cite{ll}) with
our results ($R_c=44$ km, $T_c=3.68$ MeV, $\tau_c=4.43$ s and
$t^{\mbox{\tiny off}}=0$ s) is satisfactory, the agreement being at
the level of~10\%.

We proceed and quantify the impact
of the improvements in the likelihood
described in Sect.~\ref{sec:l}.\\
The inclusion
of the detection efficiency  $\eta(E_{e^+})$
gives the single most important effect. In fact we obtain:
\begin{equation}
R_c=30\mbox{ km}, T_c=4.21\mbox{ MeV}, \tau_c=3.88\mbox{ s}.
\end{equation}
Using the new cross section $\sigma_{\bar\nu_e p}(E_{\nu_e})$
for the inverse beta decay we get the new best-fit point values:
\begin{equation}
R_c=26\mbox{ km}, T_c=4.58\mbox{ MeV}, \tau_c=3.72\mbox{ s}.
\end{equation}
Our assumption on the background $B_i$ has a
little effect on these values that become
$R_c=26$~km, $T_c=4.59$~MeV and $\tau_c=3.81$~s. \\
Similarly the inclusion of the event directions,
$\cos\theta_i$ (Eq.~(20) in \cite{sss}), that produces:
\begin{equation}
R_c=26\mbox{ km}, T_c=4.47\mbox{ MeV}, \tau_c=3.88\mbox{ s}.
\label{1best}
\end{equation}
The values of $t^{\mbox{\tiny off}}$ parameters are
always zero in these models,
and it is easy to convince oneself that this
is due to the fact that, in this one component model,
the signal is forced to decrease with the time.

The larger change concerns the radius $R_c$ that diminishes by
35\% in comparison to the value given by Lamb and Loredo,
approaching the expected neutron star radius, but still twice
larger than a typical value. In Fig.~\ref{fig0} we show the
contour plots in the $R_c-T_c$ plane, the two parameters that show
the largest correlation among them. In gray we draw the $68\%$ and
$90\%$ contours level that we obtain adopting the same likelihood
function of LL, in black the contours level when we construct the
likelihood function following Sec.~\ref{likimpr}, i.e., including
all the structural improvements discussed above.
%

With the best-fit points it is possible to estimate the total energy emitted by neutrinos in this model. Hypothesizing that in the cooling phase all
types of neutrinos are emitted, each one
carrying away an equal amount of energy, the total energy is:
\begin{equation}
\frac{E_c}{10^{53}\rm erg}=
3.39\times 10^{-6} \int^{\infty}_0
dt  \left(\frac{R_c}{\rm km}\right)^2  \left( \frac{T_c(t)}{\rm MeV}\right)^4,
\end{equation}
This value should be comparable with the gravitational
binding energy of the new born neutron star,
$E_b=(1-5)\cdot 10^{53}$~ erg.
The LL result for exponential cooling
model is $E_b=5.02\cdot 10^{53}$~erg,
namely a binding energy at the upper limit of this range,
whereas our results $E_b=3.55\cdot 10^{53}$~erg
is included in the expected range.

\subsection{Two components model\label{2cm}}
Now we include both emission phases. Following the order of
Sect.~III, we improve, step by step, the emission model of the
accretion phase. To describe accretion we add 3 new physical
parameters, namely: the positrons temperature $T_a$, the duration
of the accretion phase $\tau_a$ and the initial accreting mass
$M_a$. So the likelihood is a function of 9 parameters. The best
fit results are given in table~\ref{tab2} and will be commented in
details below.

The last column of Tab.~\ref{tab2}
shows the values $\Delta\chi^2=\chi^2_c-\chi^2_m$,
where $\chi^2_c$ is calculated with the best one-component
model of the previous section, i.e., the last case of previous
subsection (eq.~\ref{1best}), and
$\chi^2_m$ is calculated with the model described in the section
indicated in each line of Tab.~\ref{tab2}.
The larger the difference, the larger the evidence
for accretion;
a quantitative evaluation of the evidence,
taking into account the increased number of parameters, is provided later.

\subsubsection{Effect of the new likelihood--improvements
of Sect.~\ref{sec:l}\label{llstar}} The first line  of
Tab.~\ref{tab2} shows the best-fit results obtained using the
likelihood function constructed in our Sect.~\ref{sec:l} and one
of the emission model used by LL, namely the model called
``Exponential cooling and truncated accretion" (later called ECTA)
\cite{ll}. We see that the best-fit value for the initial
accreting mass $M_a$ is very large and hardly acceptable on
physical basis: this parameter is restricted by $M_a<0.6 M_\odot$
for the reasons mentioned in Sect.~\ref{am}. This result, however,
is in agreement with what found by Lamb and Loredo, who fixed
$M_a\equiv 0.5\ M_\odot$ in all models of their table V, namely
they inserted a ``prior" in their analysis. This assumption
reduces to 8 the number of free parameters and produces these
best-fit values:
\begin{equation}\label{LL}
\begin{array}{ccc}
R_c=12 \mbox{ km}, &  T_c=5.40 \mbox{ MeV}, & \tau_c=4.40\mbox{ s},\\
M_a\equiv 0.5\ M_\odot, &  T_a=2.02 \mbox{ MeV}, & \tau_a=0.70\mbox{ s}
\end{array}
\end{equation}
with $\Delta\chi^2=13.4$. We call this best-fit point $LL^*$ to
distinguish it by the true global maximum of likelihood function,
shown in Tab.~\ref{tab2}. In passing, we note that our
modifications of the likelihood do not really change the
conclusions of Lamb and Loredo \cite{previo}.

\begin{table}
\begin{center}
\begin{tabular}{|c|c|c|c|c|c|c|c|c|}
\hline
Sect. & $R_c$ & $T_c$ & $\tau_c$ & $M_a$ & $T_a$ & $\tau_a$ & $t^{\mbox{\tiny off}}$ & $\Delta \chi^2$\\
 & [km] & [MeV] & [s] & [$M_\odot$] & [MeV] & [s] & [s] & {}\\
\hline\hline
\ref{sec:l}      & 12 & 5.46 & 4.25 & 5.59 & 1.52 & 0.72 & 0. & 14.7\\
\ref{fluka}      & 14 & 4.99 & 4.76 & 0.82 & 1.75 & 0.67 & 0. & 11.2\\
\ref{shift}      & 14 & 4.88 & 4.72 & 0.14 & 2.37 & 0.58 & 0. &  9.8\\
\ref{osc}        & 16 & 4.62 & 4.65 & 0.22 & 2.35 & 0.55 & 0. &  9.8\\
\hline
\end{tabular}
\end{center}
\caption{\em The best-fit values of the astrophysical parameters for
  two components model neutrino emission. Each line of this table is
  an incremental step toward the final improved parameterization.
The last column shows the difference between the $\chi^2$
of our one-component (cooling)
model and the $\chi^2$ of each two-component model. \label{tab2}}
\end{table}

\subsubsection{New spectrum and $T_a(t)$--improvements of Sect.~\ref{fluka}}
The second line of table \ref{tab2}
shows that there is less need of the {\em a priori} on $M_a$
when we exploit the correct parametrization of the accretion
flux (accounting for the right kinematics of $e^+n$ process,
using the new cross section and allowing the positron temperature to
increase with the time). In fact the best fit value for
the initial accreting mass decreases to $M_a=0.82\ M_\odot$ in this model,
a value that is a bit larger than the reference
outer core mass, 0.6~$M_\odot$, but now closer to the expected range.
The accretion phase is characterized by a low mean energy
and by a duration shorter than a second, as expected.
The total energy carried in each phase is:
$E_c=2.0 \cdot 10^{53}$~erg and $E_a=5.7\cdot 10^{52}$~erg,
and the total binding energy is $E_b=E_a+E_c=2.5\cdot 10^{53}$~erg.
It is important to note that, in this model and as in the LL model
(see Sect.~\ref{shift})
the two emission phases are contemporaneous.
This implies that the emission spectrum is `composite',
a features that makes it easier to account for the difference
between the average energies of IMB and of
Kamiokande-II, as can be seen from  the
relatively high value of $\Delta\chi^2$ in table \ref{tab2}.

\subsubsection{Separating
accretion and cooling--improvements of
Sect.~\ref{shift}\label{shitt}} In Sect.~\ref{shift} we discussed
a procedure (`time shift') to separate temporally the accretion
and cooling phases. The time shift produces two families of best
fit values: when the IMB data fall in the accretion phase
($t^{\mbox{\tiny off}}_{\mbox{\tiny IMB}}\sim 0$) the positron
temperature $T_a$ has to increase; when  the IMB data fall in the
cooling phase ($t^{\mbox{\tiny off}}_{\mbox{\tiny IMB}}\sim
\tau_a$) the temperature $T_a$ can remain relatively low. We show
in Fig.~\ref{fig1} the likelihood profile of $T_a$ parameter,
where the above situation becomes evident.\footnote{In technical
terms, the case when  there are multiple maxima of similar quality
is termed as pathological likelihood. The statistical concept of
`pathological' solution should be distinguished from the concept
of `unphysical' solution; e.g., the one corresponding to dotted
line of Fig.~\ref{fig1} and explained in Sect.~\ref{shift}.} The
figure refers to final model that includes also neutrino
oscillations, but the same structure arises already as soon as the
time-shift is included. The other local maximum of the likelihood
besides the one shown in Tab.~\ref{tab2} is at:
\begin{equation}
\begin{array}{ccc}
R_c=10 \mbox{ km}, &  T_c=5.28 \mbox{ MeV}, & \tau_c=4.74\mbox{ s},\\
M_a= 1.27\ M_\odot, &  T_a=1.65 \mbox{ MeV}, & \tau_a=1.4\mbox{ s}
\label{altra}
\end{array}
\end{equation}
with $t^{\mbox{\tiny off}}_{\mbox{\tiny IMB}}=0.93$ s
and $\Delta\chi^2=10.2$.
The $\chi^2$ values obtained with this best-fit solution is very near
to best-fit value shown in Tab.~\ref{tab2}, the difference being
only $\sim -0.5$. Therefore this solution cannot be discarded on
statistical basis. Thus, let us examine the physical content
of this solution.
When the data of IMB belong to cooling phase, as in this solution, the
value of $T_a$ diminishes to account for the mean energy of the
first KII events, and the initial accreting mass $M_a$
increases to achieve the right number of detected events.
This implies that the family of solution
with a temporal shift
different from zero has larger values of $M_a$.
The solution of Eq.~\ref{altra} has a time constant
$\tau_a$ two-three times larger
that the expectations and, more importantly, it
has a value $M_a$ twice the outer core mass;
instead, the solution of Tab.~\ref{tab2} has a completely acceptable
value of $M_a$. Since we expect that
only a fraction of the outer core mass is exposed to
the thermal positron flux, we are led to believe
that the latter solution is more plausible than the former.

\subsubsection{Effects of oscillations--improvements
of Sect.~\ref{osc}\label{osci}}
In the last line of Tab.~\ref{tab2}, we complete the parameterization
of the flux by including neutrino oscillations.
The solution is very similar to the one described in the previous
section and the astrophysical parameters are rather similar
to the expectations.
The total energy emitted by neutrinos in
each emission phase is:
$E_c=1.8\cdot 10^{53}$ erg and
$E_a=4.8\cdot 10^{52}$~erg and the total binding
energy is $E_b=2.2\cdot 10^{53}$~erg.
The best-fit values have been obtained for normal hierarchy
and with the assumptions discussed in Sect.~\ref{osc}.
The flux of $\bar{\nu}_e$ that reaches the
detectors is a combination of the $\bar{\nu}_e$ and $\bar{\nu}_{\mu}$
emitted within the star. The best-fit values showed in the
table refers to the radius and temperature of
$\bar{\nu}_e$, that however are closely related
to the $\bar{\nu}_{\mu}$ values. In fact
the temperature of emission
is $T_c(\bar{\nu}_{\mu})=5.5$ MeV (due to Eq.~\ref{defa}) and the
radius of $\bar{\nu}_{\mu}$ neutrinosphere is
$R_c({\bar{\nu}_{\mu}})=10$~km (due to equipartition).

\begin{figure}[t]
$$\includegraphics[width=0.48\textwidth,
height=0.41\textwidth]{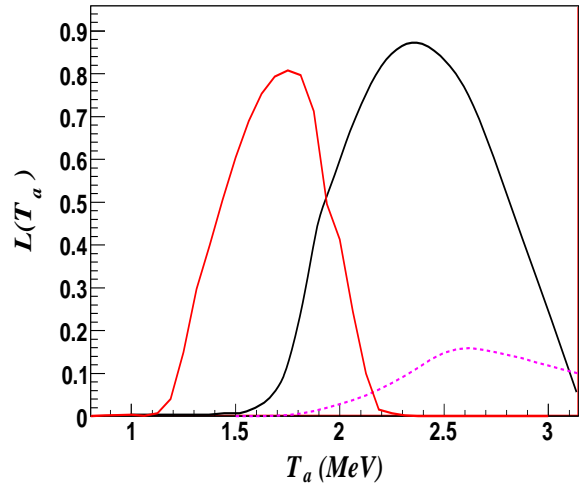}$$ \vskip-4mm \caption{\em
Profiles likelihood for $T_a$ parameter of accretion with the
complete emission model (III.C). The red line is the family of
solutions with $t^{\mbox{\tiny off}}_{\mbox{\tiny IMB}}\simeq 0.5$
s. The dark line is the other family of solutions with
$t^{\mbox{\tiny off}}=0$ and $M_a<1M_\odot$. The dotted line is
the unphysical family of solutions discussed in Sect.~\ref{shift}
with $\tau_a=0.2\ s$. \label{fig1}}
\end{figure}

The assumption that we can neglect the
$\nu_{\mu}$ during accretion implies that
the accretion flux is suppressed by the factor
$P=\cos^2\theta_{12}\sim 0.7$,
so the best-fit of initial accreting mass has to increase
by $1/P$ to maintain the same fit.
This quantitative change
has some impact on the interpretation of the
multiple solutions: In fact, the family of solutions
with $t^{\mbox{\tiny off}}_{\mbox{\tiny IMB}}\neq 0 $~ s  and with
lower values $T_a$ (red line of Fig.~\ref{fig1})
turns out to be characterized by values of $M_a$
greater than $1\ M_{\odot}$.
Thus, physical considerations on the meaning of the accreting mass
suggest, as more plausible, the family of solutions shown
with a dark line in Fig.~\ref{fig1} and in the
last line of Tab.~\ref{tab2}.

For inverted mass hierarchy, the effects of $\nu-\nu$ interactions
are known to be relevant but their quantitative treatment is still
under discussion, see e.g., \cite{simpl5,simpl6}. This
consideration, already, prevents us to draw firm conclusions.
However, adopting the formulae given in Sect.~\ref{osc} to
illustrate which are the possible effects, we can distinguish 2
main cases: large values of $\theta_{13}$, namely
$\theta_{13}>0.5^\circ$ and small values of $\theta_{13}$, namely
$\theta_{13}<0.1^\circ$. In the first case, $P_f\sim 0$ and the
survival probability $P=U_{e1}^2$; thus, we find the same results
as for normal hierarchy. In the second case, the flip probability
is $P_f\sim 1$ and $P=U_{e3}^2$ in Eq.~\ref{inverted}. The
suppression of the $\bar\nu_e$ survival probability $P\sim 0$,
along with the assumption that the flux of $\bar\nu_\mu$ and
$\bar\nu_\tau$ are very small during accretion, implies the {\em
absence} of electron antineutrino events during
accretion\footnote{Note that neglecting $\nu-\nu$ interaction, the
condition $P\sim 0$ turns out to be realized in the other case,
namely for {\em large} values of $\theta_{13}$ \cite{previo}.}.
Such a hypothesis can be tested with--and, to some extent,
excluded by--SN1987A observations \cite{previo}. All in all, the
result for inverted hierarchy should be taken with great caution
for it depends crucially not only on the present understanding of
the effects of $\nu-\nu$ interactions but also on the incomplete
description of the flux of $\bar\nu_\mu$ and $\bar\nu_\tau$ during
accretion.

\subsubsection{Remarks on the two component models\label{summat}}

Having calculated  the best fit values of the two component model,
and having understood their meaning, we pass to discuss: (a)~the
evidence for the accretion phase (namely, we compare the two
component and the one component model); (b)~the errors on the best
fit parameters for the improved model; (c)~the difference between
the best-fit ECTA model and our improved model.

\paragraph{Evidence for the phase of accretion}

In order to test whether the $\Delta \chi^2$
of the models with accretion (Tab.~\ref{tab2})
are just an effect of fluctuations,
we perform a standard likelihood ratio significance
test\footnote{We calculate
$\alpha=\prod_{i=1}^\nu \int
\exp(-x^2/2)/\sqrt{2\pi} dx_i $
for $\sum_{i=1}^\nu x_i^2 >\Delta \chi^2$ and taking
$\Delta\chi^2$ from Tab.~\ref{tab2},
where the number of random variables $x_i$ equals the
new degrees of freedom $\nu=3$.}~\cite{Cowan}.
For the four models of table \ref{tab2}, we find that we can reject
the null hypothesis (=no accretion) with a significance level of
$\alpha=0.2\%$, $\alpha=1.1\%$, and $\alpha=2.0\%$
(for the last two models) respectively
Three remarks are in order:\\
(1)~The result $\alpha=0.2\%$ basically agrees with the claim of
Lamb and Loredo: the ECTA model for emission, if correct, would
lead to an important evidence for accretion. Our improvements in
the likelihood (described in Sect.~\ref{sec:l})
do not change this inference significantly.\\
(2)~Also the other models permit to exclude the `null hypothesis'
that we test (namely, the absence of an accretion phase)
with the conventional 5\% criterion. But the
evidence becomes a bit weaker and if one prefers to be very
conservative, this could suggest caution.
This outcome can be easily understood by the
fact that our model for accretion is more constrained and can
account for certain features of the data (such as the difference of IMB
and KII energies) only at the price of some tension,
that is reflected by the increased value of $\alpha$.\\
(3)~Obviously, even the conservative attitude
does not forbid us to use the SN1987A
data to learn something on accretion. It is the question that we
formulate that changes: if we {\em assume} that the accretion phase
exists, we can ask the data to determine the model parameters.

\paragraph{Errors on the parameters}

\begin{figure}[t]
$$\includegraphics[width=0.48\textwidth,
height=0.41\textwidth]{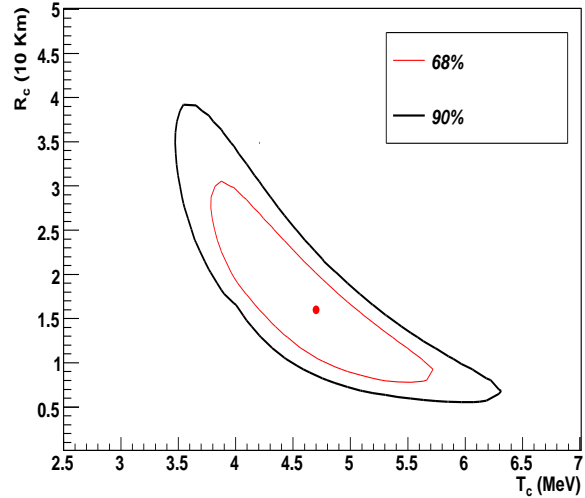}$$ \vskip-4mm \caption{\em Two
dimensional confidence regions for cooling parameters $R_c$ and
$T_c$ of $\bar{\nu}_e$ with the complete emission model
(\ref{osc}). \label{fig2}}
\end{figure}

\begin{figure}[t]
$$\includegraphics[width=0.435\textwidth,
height=0.39\textwidth]{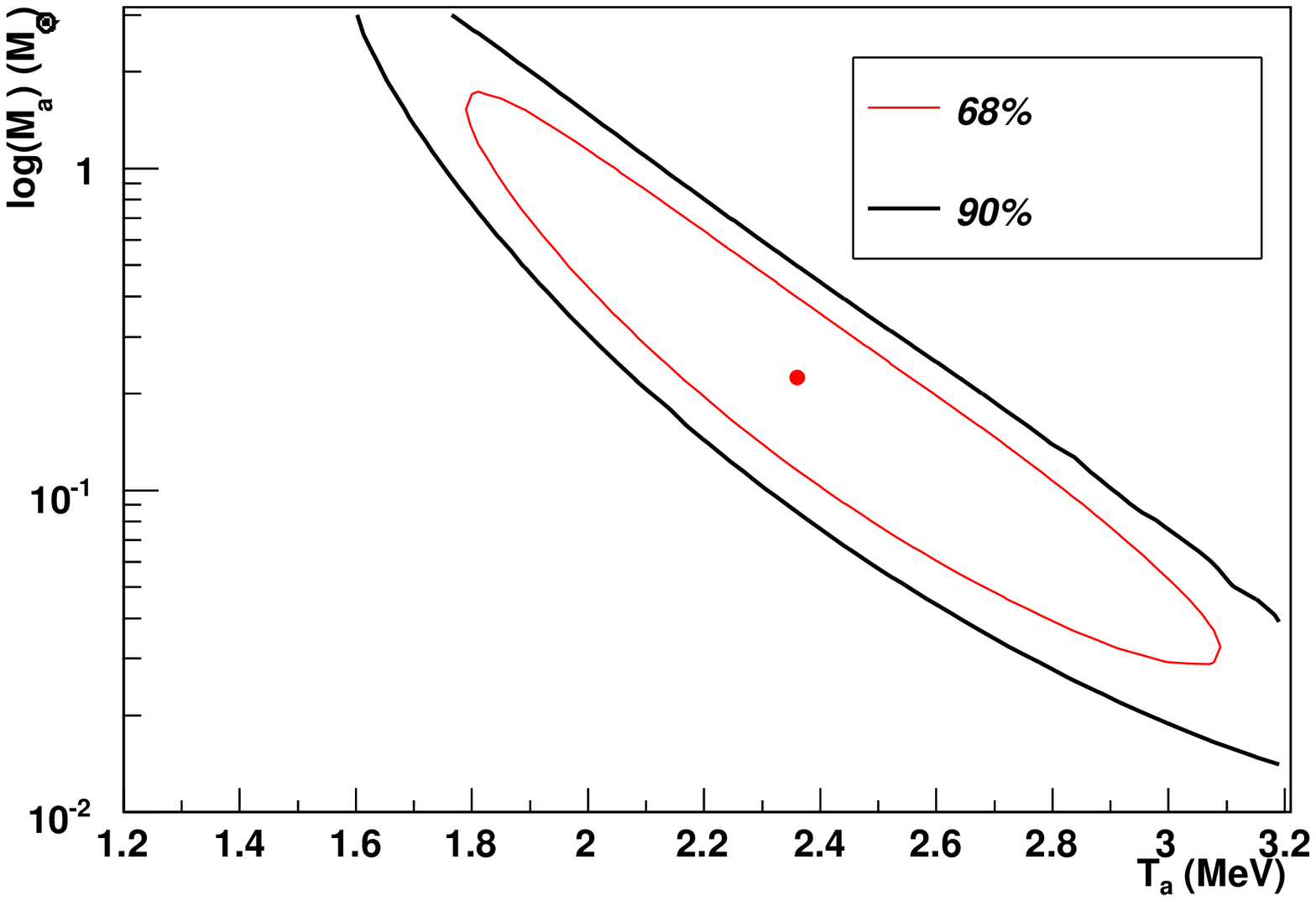}$$ \vskip-4mm \caption{\em Two
dimensional confidence regions for accretion parameters $M_a$ and
$T_a$ of $\bar{\nu}_e$ with the complete emission model
(\ref{osc}). \label{fig3}}
\end{figure}

The 1$\sigma$ errors obtained by a conventional, $\Delta \chi^2=1$,
Gaussian procedure \cite{Cowan} are:
\begin{equation}
\begin{array}{ll}
R_c=16^{+9}_{-5} \mbox{ km},         &  M_a=0.22^{+0.68}_{-0.15}\ M_{\odot},\\
T_c=4.6^{+0.7}_{-0.6}\mbox{ MeV},    &  T_a=2.4^{+0.6}_{-0.4} \mbox{ MeV}, \\
\tau_c=4.7^{+1.7}_{-1.2}\mbox{ s},   &
\tau_a=0.55^{+0.58}_{-0.17}\mbox{ s}. \label{errori}
\end{array}
\end{equation}
The 1 sided, $1\sigma$ errors for the offset times (obtained by
integrating the normalized likelihood profile~\cite{Cowan})
are:\footnote{Only IMB had a reliable measurement of the absolute
times; thus, we can use the time of its first event along with
$t^{\mbox{\tiny off}}_{\mbox{\tiny IMB}}$ to infer the moment of
the beginning of the collapse, presumably coincident with the
emission of an intense gravitational wave.}

\begin{equation}
t^{\mbox{\tiny off}}_{{\mbox{\tiny KII}}}=0.^{+0.07}\mbox{ s} ,
t^{\mbox{\tiny off}}_{\mbox{\tiny IMB}}=0.^{+0.76}\mbox{ s},
t^{\mbox{\tiny off}}_{{\mbox{\tiny BAK}}}=0.^{+0.23}\mbox{ s}.
\label{tmpop}
\end{equation}
The couples of parameters that are more tightly correlated between
them are $T_c$ with $R_c$, and $M_a$ with $T_a$. In
Fig.~\ref{fig2} and Fig.~\ref{fig3} we report the two dimensional
confidence regions for these couples of parameters showing the
$90\%$ and $68\%$ contour levels; the correlations are quite
evident from the figures. In these figures we focused on the
family of solutions with
 $t^{\mbox{\tiny off}}= 0$ after testing that the
other maxima have a very large value of accreting mass,
as discussed in Sect.~\ref{osci}.

\begin{table}[t]
\begin{center}
\begin{tabular}{|ccc||c|c|}
\hline
event  & $\delta t_i$ [s] & $E_i$ [MeV]  & $LL^*$ & {\ref{osc}} \\
\hline
K1& $\equiv 0.0$ & 20.0  & $0.69$  & $1.00$  \\
K2  & 0.107 & 13.5       & $0.92$  & $1.00$\\
K3  & 0.303 & 7.5        & $0.93$  & $0.81$    \\
K4 & 0.324 & 9.2         & $0.95$  & $0.93$ \\
K5  & 0.507 & 12.8       & $0.89$  & $0.84$\\
K6   & 0.686 & 6.3       & $0.66$      & $0.10$   \\
\hline
I1  & $\equiv 0.0$  & 38  & $0.02$  & $1.00$  \\
I2  & 0.412         & 37  & $0.02$  & $0.55$  \\
I3  &  0.650         & 28  & $0.05$ & $0.48$  \\
\hline
B1& $\equiv 0.0$ & 12.0  & 0.94   & $0.99$  \\
B2  & 0.435      & 17.9  & 0.70   & $0.85$\\
\hline
\end{tabular}
\end{center}
\caption{\em Accretion probabilities for the events occurred in the
first second; all the other events have a
probability to be due to accretion lower than 5\%.
The first three columns identify the individual events.
The last two columns are the probabilities that an event is
due to accretion. The model used in the fourth column includes
the improvements of Sect.~\ref{likimpr} and following LL sets
$M_a=0.5\ M_\odot$; the one used in the fifth column
includes also the improvements of Sect.~\ref{osc}.
\label{tabp2}}
\end{table}

\paragraph{Stability of the best-fit values}
In order to show the stability of our result we investigated:\\
1) different values for the exponent $m=1$, 3, 4 (rather than
$m=2$) in Eq.~\ref{tap}, that describes the temporal
behavior of positron temperature during accretion;\\
2) the value $k=10$ (rather than $k=2$) in Eq.~\ref{jcut} that
describes the sharpness of the transition between accretion and
cooling phases;\\
3) deviations from the hypothesis of equipartition, by increasing
or decreasing the ratio between the $\bar\nu_e$ and $\bar\nu_x$
luminosities by a factor of 2 \cite{keil}.\\
In all cases, the  $\chi^2$ changes less than 1 with respect to
the best fit result; furthermore, the best fit values of the
astrophysical parameters change only within their 1 $\sigma$
errors given in Eq.~\ref{errori}.

\paragraph{On the differences with the parameterization by Lamb and Loredo}

\begin{figure}[t]
\includegraphics[width=0.21\textwidth,height=0.21\textwidth]{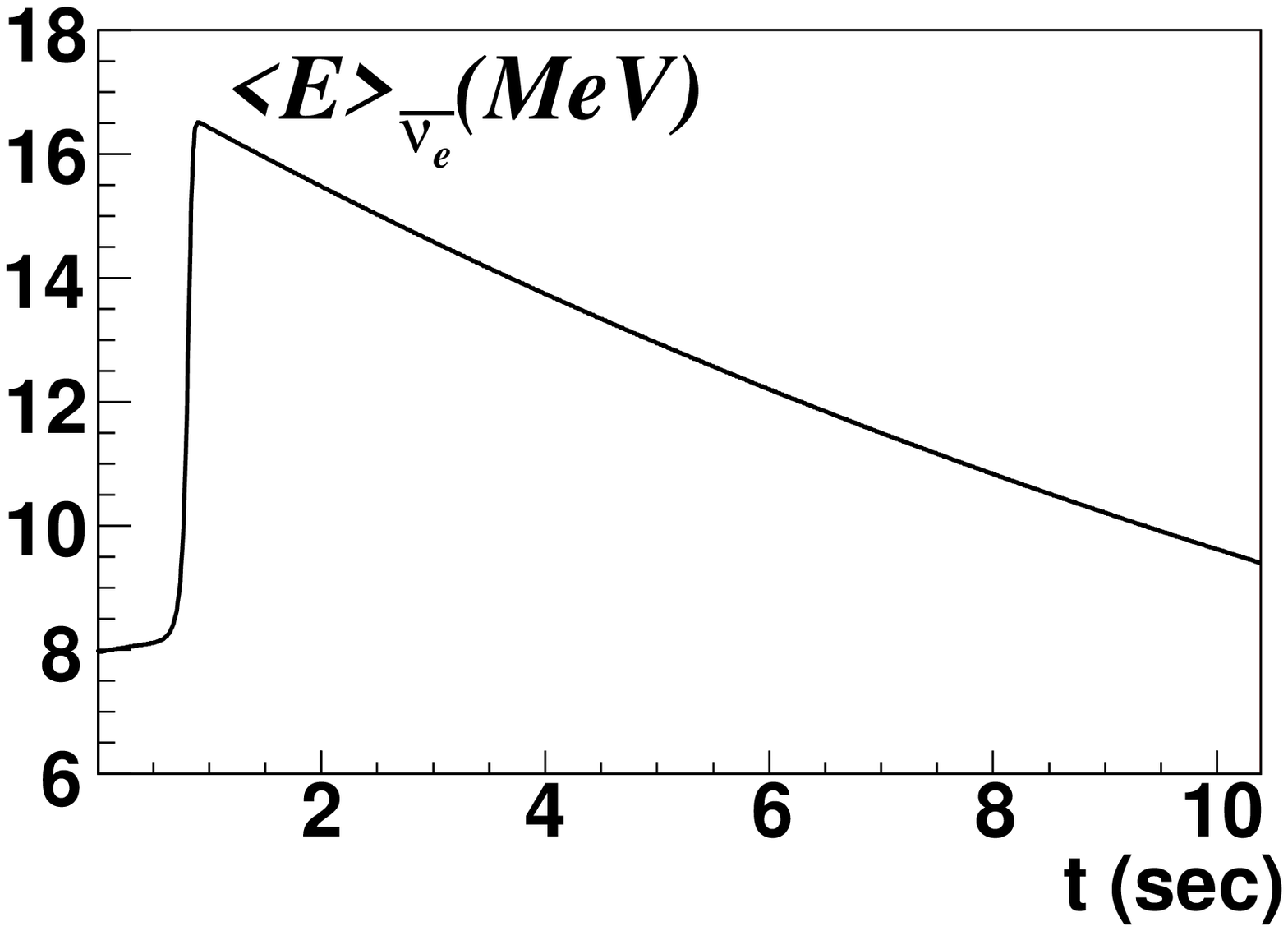}
\includegraphics[width=0.21\textwidth,height=0.21\textwidth]{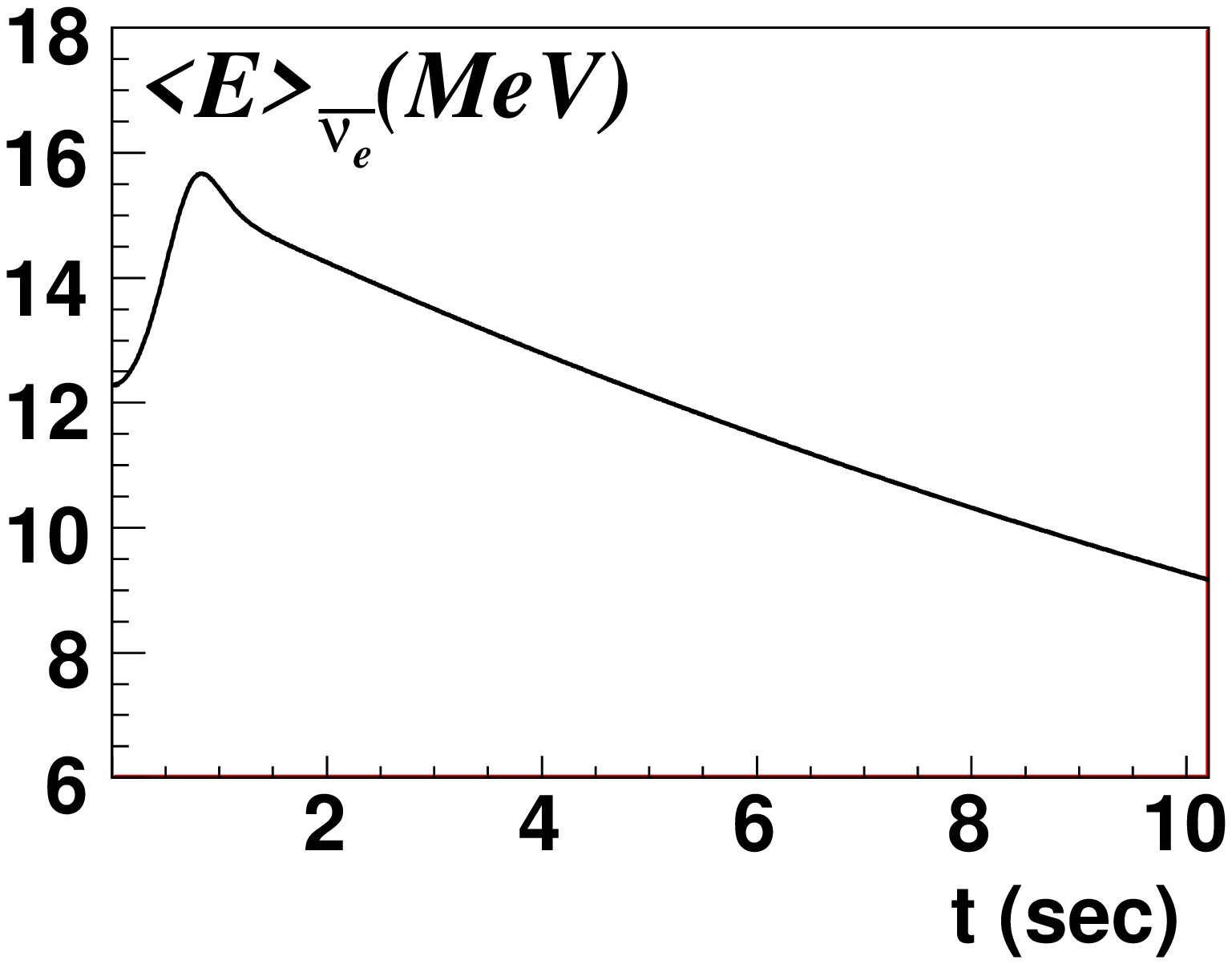}
\vskip-4mm \caption{\em $\bar{\nu}_e$ mean energy as a function of
the time in the $LL$ ECTA model (left panel) and in the \ref{osc}
model (right panel).\label{fig4}}
\end{figure}

\begin{figure}[t]
\includegraphics[width=0.21\textwidth,height=0.21\textwidth]{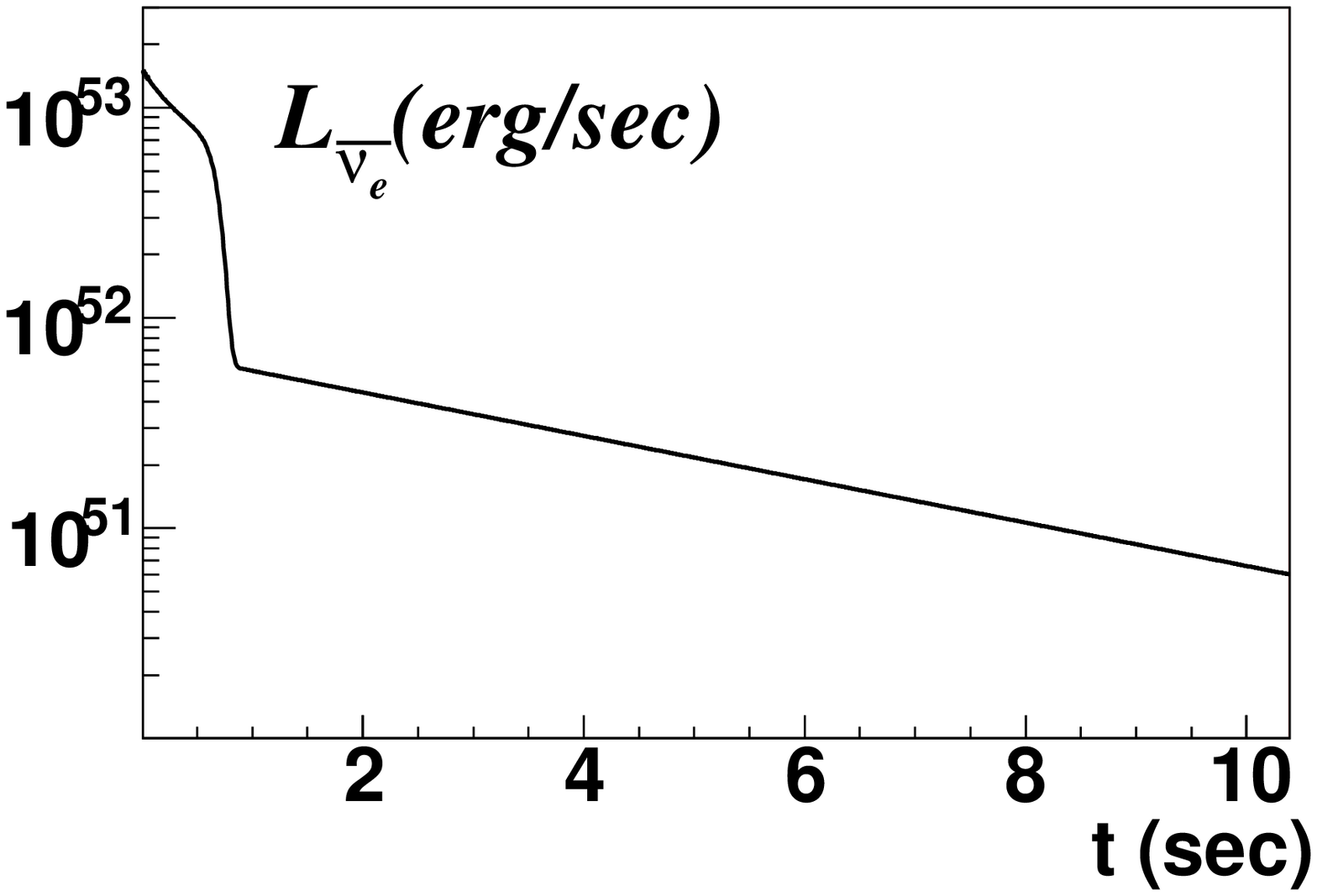}
\includegraphics[width=0.21\textwidth,height=0.21\textwidth]{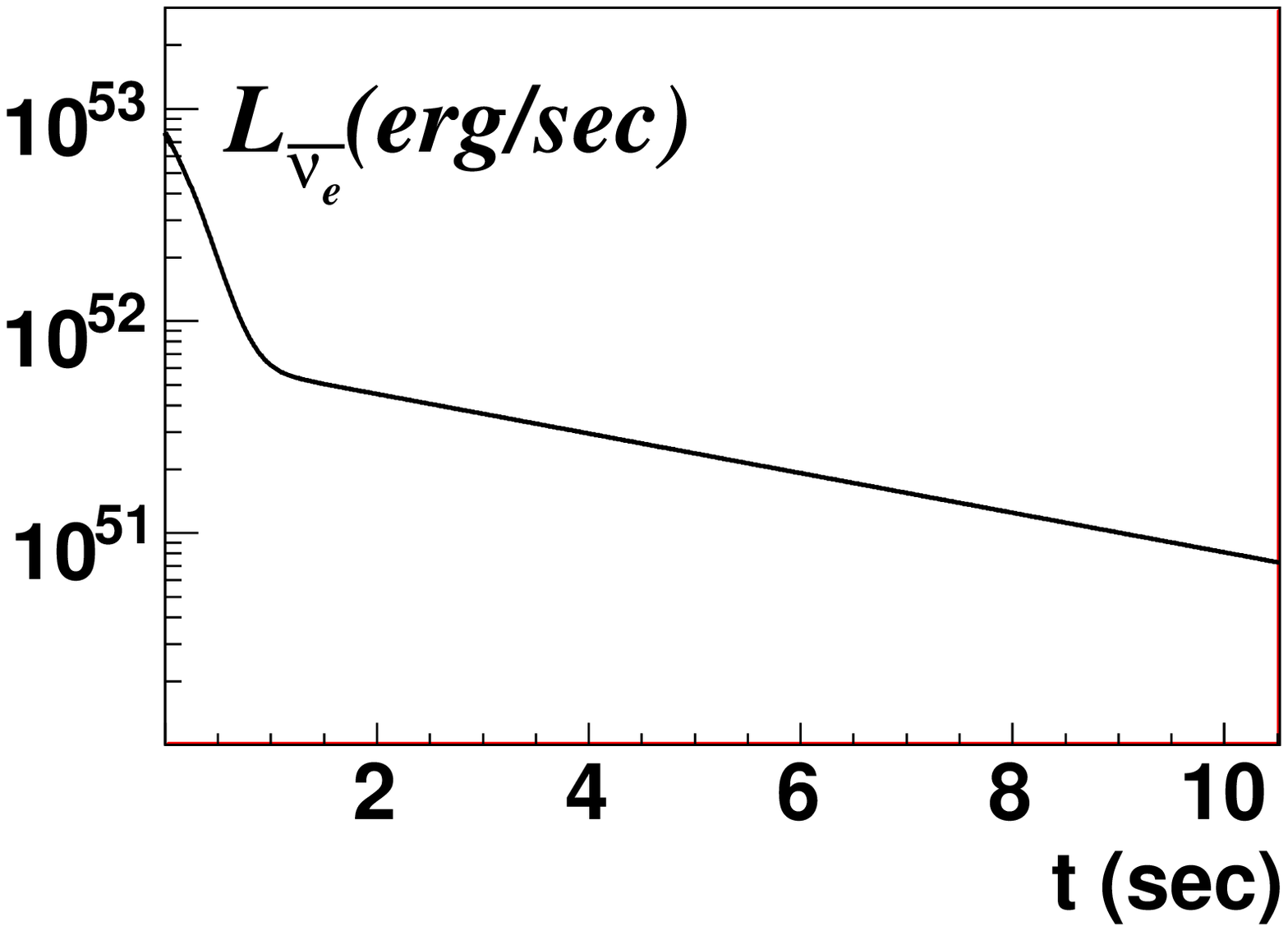}
\vskip-4mm \caption{\em $\bar{\nu}_e$ luminosity in the $LL$ ECTA
model (left panel) and in the \ref{osc} model (right panel).
\label{fig5}}
\end{figure}

To illustrate better the difference between the ECTA model and our
final emission model, we show in Tab.~\ref{tabp2} the
probabilities of the individual events to be due to accretion. A
direct comparison with LL results \cite{ll} is not possible, since
a similar table is not given there; thus, we repeat their
calculation following their prescriptions and make reference to
the model $LL^*$ described in Sect.~\ref{llstar}. The column
$LL^*$ shows that the early KII and Baksan events are due to
accretion and those by IMB to cooling. This proves that, assuming
the Lamb and Loredo ECTA model, the fit takes advantage of the
fact that the assumed energy distribution is `composite': the low
energy events of KII are explained by the accretion component,
the high energy events of IMB instead by the
high energy tail due to the cooling component.\\
The results of our model are shown in the column denoted
by \ref{osc}. Since our model has (by construction) a
quasi-thermal spectrum at any time, both KII and IMB early events
have a large probability to be due to accretion. The tension
between the different energies of KII and IMB events leads to a
slightly worse $\chi^2$ (see Tab.~\ref{tab2}).\\
Finally, we compare in Fig.~\ref{fig4} the mean energy of
$\bar{\nu}_e$ in the $LL^*$ accretion model, left panel, with the
trend of the mean energy in our model \ref{osc}, right panel,
whereas in Fig.~\ref{fig5} we plot the $\bar{\nu}_e$ luminosity
obtained with the $LL^*$ model (left panel) and the model of
Sect.~\ref{osc} (right panel). The features of these curves are
similar to those found in typical numerical simulations,
see e.g.~\cite{janka new}, with the
exception  of the average energy curve of the LL model that has a
very pronounced jump.


\section{Summary}
We presented an improved analysis of observations
of SN1987A by Kamiokande-II, IMB and Baksan.
We recall the main points:

1.~We collected in Sect.~\ref{sec:l}
a large number of technical improvements:
new detection cross section, procedure to
include the information on the direction of the events,
treatments of the background and of the efficiency.
The most relevant improvement is the last one,
but none of these changes affect SN1987A data
analysis in a crucial manner.

2.~We described in Sect.~\ref{osc}
the various effects of neutrino oscillations:
the oscillations in the star (including the self interaction of neutrinos)
and the Earth matter effect.
We verified that oscillations with normal hierarchy
do not affect the results in an essential manner and discussed
the inverted hierarchy case.

3.~We proposed in Sect.~\ref{sec:m}
a new parameterization of the flux of electron
antineutrinos emitted in the accretion phase.
We improved on the energy spectrum, and most
importantly, on the time-distribution:
(a)~We prescribed the temperature of the positrons
to increase in such a manner that
at $t\sim \tau_a$ the average energy antineutrinos is
approximatively continuous, that  overcomes
a shortcoming of the parameterization of \cite{ll}
noted in \cite{rm} and recalled in the Introduction.
(b)~We also prescribed that the number of neutrons exposed to
the positron flux decreases in time more smoothly
than as in \cite{ll};
in this way the luminosity is also continuous, as expected on general basis.
(c)~Finally, we avoided the simultaneous presence of cooling and accretion
antineutrinos by time-shifting (delaying) the cooling phase of
an amount $t\sim \tau_a$, again improving on \cite{ll}.

4.~We demonstrated that the improvements on the  parameterization
are the most important. The most striking feature is that the best
fit parameters are rather similar to the expectations of the Bethe
and Wilson scenario. This is in contrast with what happens using
the Lamb and Loredo parameterizations of the flux, where it is
needed to impose a prior in the analysis to avoid best-fit values
outside the physical ranges. Furthermore, our flux leads to smooth
luminosity curves and average energies. [The most appealing
outcomes of their analysis, such as the neutrino sphere radius
resembling  the neutron star mass, the duration of the cooling
phase $\sim 0.5$ s, the total amount of emitted energy, are
practically unchanged.]

5.~We evaluated the errors and the correlations on the
parameters and we can rule out the hypothesis
of a one-phase model with a significance of  2\%.

{}From these calculations one
draws two main messages: an accurate
choice of the model for the analysis
of SN1987A observations is important;
the agreement between the observations and
the conventional expectations is more than encouraging
in the proposed model.
We hope that these results will help to progress further
in the understanding of this epochal observation.



\acknowledgments
We thank
A.~Drago, D.K.~Nadyozhin, S.~Pastor, and F.L.~Villante
for useful discussions.
The work is partly supported by the MIUR grant for the Projects of
National Interest PRIN 2006 ``Astroparticle Physics''
and by the European FP6 Network ``UniverseNet''
MRTN-CT-2006-035863.

\appendix*
\section{Accretion energy spectrum}
We derive the equations we used to describe
the energy spectrum of accretion $\bar\nu_e$.
\subsection{Derivation of Eq.~\ref{flusso}}
We assume that, during accretion, the antineutrino flux is mostly
produced by the weak reaction $e^+ n\to \bar\nu_e p$.
The rate of this reaction is given by:
\begin{equation}
\Gamma= N_n \int_{m_e}^\infty dE_{e^+} \ \frac{d\, n_{e^+}}{d\, E_{e^+}}\ \beta_e c \int dE_\nu\ \frac{d\, \sigma_{e^+n}}{d\,
E_\nu}, \label{prema}
\end{equation}
where $N_n$ is the number of target neutrons, assumed
to be at rest, and
where $\beta_e$
is the positron velocity in
natural units.
The second integral yields the cross section as a
function of the positron energy $E_{e^+}$.
The distribution of the positrons
$d\, n_{e^+}=2 d^3 p_e/h^3 /(1+\exp[(E_{e^+}-\mu_{e^+})/T_a])$
has a negligible chemical potential \cite{janka}, thus
\begin{equation}
\frac{d\, n_{e^+}}{d\, E_{e^+}}(E_{e^+})= \frac{8 \pi \beta_e}{(h c)^3}\  g_{e^+}(E_{e^+},T_a),
\end{equation}
where $g_{e^+}$ is given by Eq.~\ref{fluacc}.
The range of
integration of $E_\nu$ in Eq.~\ref{prema} can be easily found
from the expression:
\begin{equation}
E_\nu=\frac{E_e+\delta_+}{1+(E_e-p_e \cos\phi)/{m_n}},
\end{equation}
where $\cos\phi=\hat{n}_e\cdot \hat{n}_\nu$ is the cosine of the angle between the directions of the positron and the antineutrino,
$\cos\phi\in[-1,1]$, and where $\delta_+=(m_n^2-m_p^2+m_e^2)/(2 m_n)=
1.293\mbox{ MeV}\approx \delta_-$ in Eq.~\ref{kinem}.
The dependence on the antineutrino energy $E_\nu$ is as follows:
\begin{equation}
\frac{d\, \Gamma}{d E_\nu}= \frac{8\pi c}{(h c)^3} N_n \int d E_{e^+}\ \beta_e^2 \ g_{e^+} \ \frac{d\, \sigma_{e^+n}}{d\,
E_\nu}. \label{corro}
\end{equation}
The cosine can take all values
for fixed $E_\nu\ge E_{min}$, with
\begin{equation}
E_{min} = \frac{m_e+\delta_+}{1+m_e/m_n}\approx 1.803\ \mbox{MeV},
\end{equation}
[for smaller $E_\nu$, only
certain values of $\cos\phi$ around $\cos\phi=-1$ are allowed; but this
happens in an interval of $E_\nu$ wide only about 1 eV].
The range for $E_{e^+}$ is:
\begin{equation}
E_{e^+}=
{\textstyle
\frac{(E_\nu-\delta_+) (1-\epsilon)-\epsilon \cos\phi
  \sqrt{(E_\nu-\delta_+)^2- m_e^2 (1+\Delta)}
}{1+\Delta},}
\label{narro}
\end{equation}
where $\epsilon={E_\nu}/{m_n}$ and
$1+\Delta=(1-\epsilon)^2-\epsilon^2 \cos\phi^2$.
From this equation it is clear that the $E_{e^+}$ range is pretty
narrow for the energies $E_\nu\ll m_n$ in which we are interested;
thus, we approximate the expression in Eq.~\ref{corro} as:
\begin{equation}
{\textstyle
\frac{d\, \Gamma}{d\, E_\nu} \approx \frac{8\pi c}{(h c)^3} N_n
g_{e^+}( \bar{E}_{e^+}(E_\nu), T_a) \int\! d E_{e^+}\! \beta_e^2 \
\frac{d\, \sigma_{e^+n}}{d\, E_\nu}} \label{edef}
\label{rarra}
\end{equation}
where the positron distribution is calculated at the central point of the interval of cosine, namely at $\cos\phi=0$:
\begin{equation}
\bar{E}_{e^+}(E_\nu)=\frac{E_\nu-\delta_+}{1-E_\nu/m_n}. \label{emed}
\end{equation}
Eq.~\ref{rarra} gives the $\bar\nu_e$ flux in
Eq.~\ref{flusso}; the integral will be discussed a while.
The advantage of Eq.~\ref{edef} is that the dependence  on the
parameter $T_a$ has been extracted from the integral. The integral can be
calculated once forever and we are left with a simpler expression. We
checked that, in the most relevant range $T_a=1-4$ MeV, the
approximated expression agrees with the correct one at the 1\% level
for energies $E_\nu<10\, T_a$, and even better when $d \Gamma/d E_\nu$ is
integrated in $E_\nu$: indeed, the rate $\Gamma$ is precise at 0.1\%.


\subsection{Derivation of Eq.~\ref{appra}\label{laterio}}
The cross section in Eq.~\ref{appra}, precisely defined as a numerical
approximation of the integral in Eq.~\ref{rarra},
was obtained adapting
the calculation of \cite{sss} for the inverse beta decay reaction.
In fact, the differential cross section
\begin{equation}
\frac{d\, \sigma_{e^+n}}{d\, E_\nu}(E_{e^+},E_\nu)  = \frac{G_F^2\ \cos^2\theta_C}{256\pi\ m_n\ p_e^2} |{\mathcal M}|^2 (1+r)
\end{equation}
has the {\em same} matrix element $|{\mathcal M}|^2(s-u,t)$.
What changes is the expression of the invariants, now given by: $s-u=2 m_n
(E_\nu+E_e)+m_e^2$ and $t=m_p^2-m_n^2+2 m_n (E_\nu-E_e)$.
The factor $r(E_e)$ describes the small QED radiative
corrections; we use expression in \cite{sss}.
The constant in front to the differential cross
section is 2 times smaller than the one for inverse beta decay,
because the antineutrino has 1 helicity state
whereas the positron has 2.
The characteristic
$1/\beta_e$ behavior
of an exothermic reaction (such as $e^+ n \to \bar\nu_e p$) is
compensated by the 2 explicit factors $\beta_e$
from the positron phase space and from the relative velocity
between $e^+$ and $n$ in the reaction
rate, included in $\sigma_{e^+n}$.

Our cross section compares well with the
approximation of Tubbs and
Schramm \cite{tubbs,monch}:
\begin{equation}
\sigma_{e^+n}^{TS}=1.7 \times 10^{-44}  \frac{1+3 g_A^2}{8} \left(\frac{E_\nu}{m_e}\right)^2,
\end{equation}
with $g_A=-1.27$, since the percentage deviation $100 (1-\sigma_{e^+n}^{TS}/\sigma_{e^+n})$ at $E_\nu=5,10,20,30$ MeV is just -1\%, -2\%, -6\%, -11\%
(or -1\%, -3\%, -7\%, -10\% when comparing with the approximation of $\sigma_{e^+n}$ in Eq.~\ref{appra}).

The cross section used in \cite{ll} is formally less correct,
since it is the same as the above approximation but replacing
$g_A^2\to |g_A|\approx 1.254$. [This is stated
in Eq.~(4.5) of LL and can be checked by the value of the energy radiated
during accretion, their Eq.~(6.2)]. The deviation $100
(1-\sigma_{e^+n}^{LL}/\sigma_{e^+n})$ is not large; for $E_\nu=5,10,20,30$ MeV
is 17\%, 16\%, 13\%, 10\% (or 18\%, 16\%, 13\%, 10\% when
comparing with Eq.~\ref{appra}).

Our
parametrization of $\bar\nu_e$ spectrum, Eq.~\ref{fluacc}, differs
also for another reason with the one of LL,
since we adopt the positron flux calculated
in $\bar{E}_e(E_\nu)$ (defined in Eq.~\ref{emed}),
whereas LL use the antineutrino flux calculated in $E_\nu$,
namely, $g\to E_\nu^2/[1+\exp(E_\nu/T_a)]$. Also this
modification acts in the
direction of increasing the expected flux. The
difference can be quantified by evaluating
the integral of the fluxes
$\Phi_{\bar\nu_e}(T_a)=\int \frac{d\Phi_{\bar\nu_e}}{d E_\nu}\, dE_\nu $:
indeed, $1- \Phi^{LL}_{\bar\nu_e}(T_a)/\Phi_{\bar\nu_e}(T_a) =$
54\%, 35\%, 26\% or 19\% for $T_a=1$, $2$, $3$ or $4$ MeV; namely, our flux is
significantly larger.

\end{document}